\UseRawInputEncoding
\documentclass[manuscript]{aastex}

\usepackage{lineno}
\usepackage{tikz}
\usepackage{amsmath}

\usepackage{graphicx}

\usepackage{pdflscape}

\slugcomment{\textbf{Revised 2021 March 17}}

\shorttitle{Comet Rotation}
\shortauthors{Jewitt}

\begin{document}

\title{Systematics  and Consequences of Comet Nucleus   \\Outgassing Torques }

\author{David Jewitt$^{1,2}$
} 
\affil{$^1$Department of Earth, Planetary and Space Sciences,
UCLA, 
595 Charles Young Drive East, 
Los Angeles, CA 90095-1567\\
$^2$Department~of Physics and Astronomy,
University of California at Los Angeles, \\
430 Portola Plaza, Box 951547,
Los Angeles, CA 90095-1547\\
}

\email{jewitt@ucla.edu}

\begin{abstract}

Anisotropic outgassing from comets  exerts a torque sufficient to rapidly change the angular momentum of the nucleus, potentially leading to rotational instability.  Here, we use empirical measures of   spin changes in  a sample of  comets to characterize the torques and to compare them with expectations from a simple model. Both the data and the model show that the characteristic spin-up timescale, $\tau_s$, is a strong function of nucleus radius, $r_n$.    Empirically, we find that the timescale  for comets (most with perihelion  1 to 2 AU and eccentricity $\sim$0.5) varies as $\tau_s \sim 100 r_n^{2}$, where $r_n$ is expressed in kilometers and $\tau_s$ is in years.   The  fraction of the nucleus surface that is active varies as $f_A \sim 0.1 r_n^{-2}$.       We find that the median value of the dimensionless moment arm of the torque is $k_T$ = 0.007 (i.e.~$\sim$0.7\% of the escaping momentum torques the nucleus), with weak ($<$3$\sigma$) evidence for a size dependence $k_T \sim 10^{-3} r_n^2$.    Sub-kilometer nuclei have spin-up timescales comparable to their orbital periods, confirming that outgassing torques are  quickly capable of driving small nuclei towards rotational disruption.   Torque-induced rotational instability likely accounts for  the paucity of sub-kilometer short-period cometary nuclei, and   for the pre-perihelion destruction of sungrazing comets.   Torques from sustained outgassing  on small active asteroids can rival YORP torques, even for very small ($\lesssim$1 g s$^{-1}$) mass loss rates.  Finally, we highlight the important role played by observational biases in the measured distributions of $\tau_s$, $f_A$ and $k_T$.
\end{abstract}

\keywords{}

\section{INTRODUCTION}
The dynamical lifetimes of short-period comets are about 0.5 million years, some $10^{-4}$ of the age of the solar system, while the physical lifetimes are at least an order of magnitude shorter still (Levison and Duncan 1997).  Several processes potentially limit the physical lifetimes, including complete devolatilization of the nucleus, formation of a global, refractory mantle that stifles outgassing, and rotational disruption from outgassing torques (Samarasinha 1986, Jewitt 1992, 1997).       Knowledge of the physical lifetimes is important both for understanding the populations of the Kuiper belt and Oort cloud source reservoirs (with shorter lifetimes requiring larger source populations in order to maintain steady-state) and for understanding the evolutionary properties of comets when in the terrestrial planet region.

Reaction forces from sublimation  exert a torque that can change both the magnitude and the direction of the  spin of a cometary nucleus (Whipple 1961).  It has long been noticed that the characteristic timescale for changing the spin can be very short (Samarasinha 1986, Jewitt 1992, 1997), and that the lifetimes of cometary nuclei when active may be determined by spin-up to the point of rotational disruption.  Originally proposed  in the near absence of relevant rotational and physical data on cometary nuclei, we now possess a better physical characterization of the comets and several reliable measurements of  nucleus spin changes that can be used to better-define the spin-up process.  

Spin-up has been discussed in the refereed literature by Jewitt (1997), Gutierrez et al.~(2003), Samarasinha and Mueller (2013),  Steckloff and Jacobson (2016), Mueller and Samarasinha (2018), Kokotanekova et al.~(2018), Rafikov (2018), and Steckloff and Samarasinha (2018).  In this paper we first recap the simple spin-up model of Jewitt (1997), then describe recent measurements to establish the empirical  nucleus spin-up timescale as a function of nucleus parameters. We then consider the consequences of this timescale for the spin evolution of outgassing cometary nuclei. Finally, we discuss the role of observational bias.

\section{SCALING RELATIONS}

A torque applied to a rotating nucleus changes the vector angular momentum, resulting in both excited (non-principal axis) rotation and a changing spin period.  Excited rotational states have been reported, perhaps most beautifully in 1P/Halley (Samarasinha and A'Hearn 1991).  However, non-principal axis rotation is generally much more difficult to detect than changes in the magnitude of the spin, manifested as time-dependence in the period deduced from lightcurve photometry (Gutierrez et al.~2003).  Therefore, we focus on the effect on the spin rate of a torque exerted by non-uniform mass loss. We write the scalar torque as

\begin{equation}
T = k_T(r_n) V_{th} r_n \overline{\dot{M}}
\label{T}
\end{equation}

\noindent where  $r_n$ is the radius, $\overline{\dot{M}}$ [kg s$^{-1}$] is the average rate of mass loss, $V_{th}$ is the speed with which material is ejected and $k_T(r_n)$ is the ``dimensionless moment arm''.   The momentum is dominated by outflowing gas and, therefore, $\overline{\dot{M}}$ and $V_{th}$ refer to the gas production rate and speed, respectively, and momentum in the dust is ignored.  Quantity $k_T$ is equal to the fraction of the outflowing momentum that exerts a torque on the nucleus.  The limiting values are $k_T$ = 0, corresponding to isotropic ejection with no net torque and $k_T$ = 1, corresponding to collimated ejection in a direction tangent to the surface.  The spin angular momentum is $L = I \omega$, with $I$ equal to the moment of inertia and  angular speed of the rotation $\omega = 2\pi/P$, where $P$ is the instantaneous rotation period.  The shapes of cometary nuclei are typically irregular and $I$ cannot be generally defined.  For simplicity, we represent the nucleus by a homogeneous sphere, for which $I = (2/5) M_n r_n^2$, where $M_n = (4/3)\pi \rho_n r_n^3$ is the nucleus mass and $\rho_n$ is the nucleus density.  Equivalently, 

\begin{equation}
L = (8\pi/15) \rho_n r_n^5 \omega.
\label{L}
\end{equation}

\noindent  Then, defining the characteristic timescale for spin-up by the torque as $\tau_S = L/T$, we obtain from Equations \ref{T} and \ref{L} (c.f.~Jewitt 1997)

\begin{equation}
\tau_s = \left(\frac{16\pi^2}{15}\right)  \left(\frac{\rho_n r_n^4}{k_T(r_n) V_{th}  P}\right) \left(\frac{1}{\overline{\dot{M}}}\right).
\label{tau_s}
\end{equation}

\noindent As noted above, $\tau_s$ and $P$  in Equation (\ref{tau_s}) can be extracted from lightcurve observations (e.g.~Kokotanekova et al.~2018).    The other parameters in Equation (\ref{tau_s}) deserve comment, as follows.  

\textit{Density, $\rho_n$:} Only the density of the nucleus of 67P/Churyumov-Gerasimenko has been directly  measured. Published values for this, and for a range of nuclei studied using less direct techniques, are compatible with $\rho_n$ = 500 kg m$^{-3}$ (Groussin et al.~2019), which we adopt here.  

\textit{Speed, $V_{th}$:} The momentum of the ejected material originates in the thermal motions of gas produced by  sublimated cometary ice.  To first order, we take the speed of the sublimated gas as the mean thermal speed, $V_{th} = (8 k T/(\pi \mu m_H))^{1/2}$, where $k = 1.38\times10^{-23}$ J K$^{-1}$ is the Boltzmann constant, $T$ is the temperature of the sublimating surface, $\mu$ is the molecular weight and $m_H = 1.67\times10^{-27}$ kg is the mass of the hydrogen atom.    Setting $\mu$ = 18 for water, the dominant cometary volatile,  and $T$ = 330 K for the hemispheric temperature at 1 AU, we obtain speed  $V_{th} \sim $ 677 m s$^{-1}$.  At 2 AU, we find $T$ = 233 K and $V_{th}$ = 522 m s$^{-1}$.   The distance dependence of the speed is weak (because $V_{th} \propto T^{1/2}$ and $T \propto r_H^{-1/2}$), a fact confirmed by high resolution spectroscopic measurements giving $V_{th} \propto r_H^{-1/4}$ over the range 1 $\le r_H \le$ 8 AU  (Biver et al.~2002). Sublimation depresses the temperature below the local blackbody value to $T$ = 205 K near $r_H$ = 1 AU, corresponding to $V_{th}$ = 490 m s$^{-1}$.  Noting the narrow range of heliocentric distances (1 $\lesssim r_H \lesssim$ 2 AU)  over which most of the comets of this study were observed, we neglect any heliocentric variation and set $V_{th}$ = 500 m s$^{-1}$, which is within a factor $\sim$2 of speeds measured in cometary gas in this distance range (Biver et al.~2002).

\textit{Mass Loss Rates, $\dot{M}$:} In the sublimation hypothesis we expect that activity should be proportional to the nucleus surface area, and write 

\begin{equation}
\overline{\dot{M}} = 4 \pi f_A(r_n)  \overline{f_s(r_H)} r_n^2,
\label{area}
\end{equation}

\noindent where $4\pi r_n^2 f_A(r_n)$ is the sublimating  area of the nucleus, assumed spherical, and $\overline{f_s(r_H)}$ is the orbitally-averaged sublimating mass flux (kg m$^{-2}$ s$^{-1}$) calculated  from the energy balance equation as described in the Appendix.  Quantity $f_A$ is known as the ``active fraction'',  equal to the ratio of the sublimating area to the surface area of a sphere having the radius $r_n$.

Combining Equations (\ref{tau_s}) and (\ref{area}), we have

\begin{equation}
\tau_s = \left(\frac{4\pi}{15}\right)  \left(\frac{\rho_n  r_n^2}{k_T(r_n) f_A(r_n) V_{th}  P}\right) \left(\frac{1}{\overline{f_s(r_H)}}\right)
\label{tau_s2}
\end{equation}

\noindent showing that we should expect the characteristic spin-up timescale to vary as $\tau_s \propto r_n^2$, but only if $f_A k_T P$ in the denominator is independent of $r_n$.  Period $P$ is measured for each nucleus in this study.  In the next section we calculate $\tau_s$, $k_T$ and $f_A$ from published data to compare with this expectation.

\section{EMPIRICAL RELATIONS}
\label{empirical}


\subsection{Spin-Up Timescale, $\tau_s$}

The first reviews of cometary nucleus rotation (Sekanina~1981, Whipple~1982) were published before useful rotation data were available and, as a result, are mainly of historical interest.  The first reliable measurements of a cometary nucleus rotational lightcurve were those of 49P/Arend-Rigaux, obtained in the mid-1980s (A'Hearn et al.~1985, Jewitt and Meech 1985).  Before that time it was widely held that the nucleus could not be directly detected in ground-based observations; the study of low activity comets like 49P/Arend-Rigaux revealed this belief to be unfounded.   However,  it remains true that rotational lightcurves can be directly determined in relatively few comets because of photometric contamination by coma.  Unlike the asteroids, comets usually exhibit a diffuse appearance due to outgassed material in the coma, resulting in the dilution  of the nucleus rotational lightcurve to unobservable levels.  In some active objects, however, periodic structures including jets and spirals in the coma can be used to infer the rotation period of the nucleus, even though the nucleus itself cannot be photometrically isolated (e.g.~Samarasinha and A'Hearn 1991).  

In this work, we use only published measurements of nucleus rotation and rotation changes, for which Kokotanekova et al.~(2018) presented a convenient summary.    These authors list (in their Table 2) the measured change in the rotation period \textit{per cometary orbit}, $|\Delta P|$,  which is related to the spin-up timescale, $\tau_s$, by 

\begin{equation}
\tau_s = \frac{P}{|\Delta P|} P_K
\label{timescale}
\end{equation}

\noindent where $P$ is the measured instantaneous rotation period of the nucleus and $P_K$ is the Keplerian orbital period ($P_K = a^{3/2}$, with $P_K$ in years and orbital semimajor axis $a$ in AU).    We add rotational measurements of 46P/Wirtanen using data from Farnham et al.~(2021), but ignored two, earlier measurements of this object by Meech et al.~(1997) and Lamy et al.~(1998) because their results were discordant yet nearly simultaneous.   Exclusion of 46P/Wirtanen from our sample would not change any of the following results. The measurements are listed in Table (\ref{sublimation}).

 Equation (\ref{timescale}) gives a measure of how long the nucleus would take to change from stationary to its present rotation period, assuming that the orbitally averaged torque is constant.  In most comets, the period drifts  slowly, and the reported period changes are noticed only when comparing determinations made in different orbits.     In comets 41P/Tuttle-Giacobini-Kresak, 46P/Wirtanen and 103P/Hartley, the  rotational period varies so quickly that the rate of change, $dP/dt$, can be measured within a single orbit (Drahus et al.~2011, Knight et al.~2015, Bodewits et al.~2018, Moulane et al.~2018, Schleicher et al.~2019, Farnham et al.~2021). Note that, while $\Delta P$ can be positive or negative, and a given nucleus can be either spinning up or spinning down,  we are interested only in the magnitude of the change, $|\Delta P|$.

\begin{figure}[ht]
\centering
\includegraphics[width=0.850\textwidth]{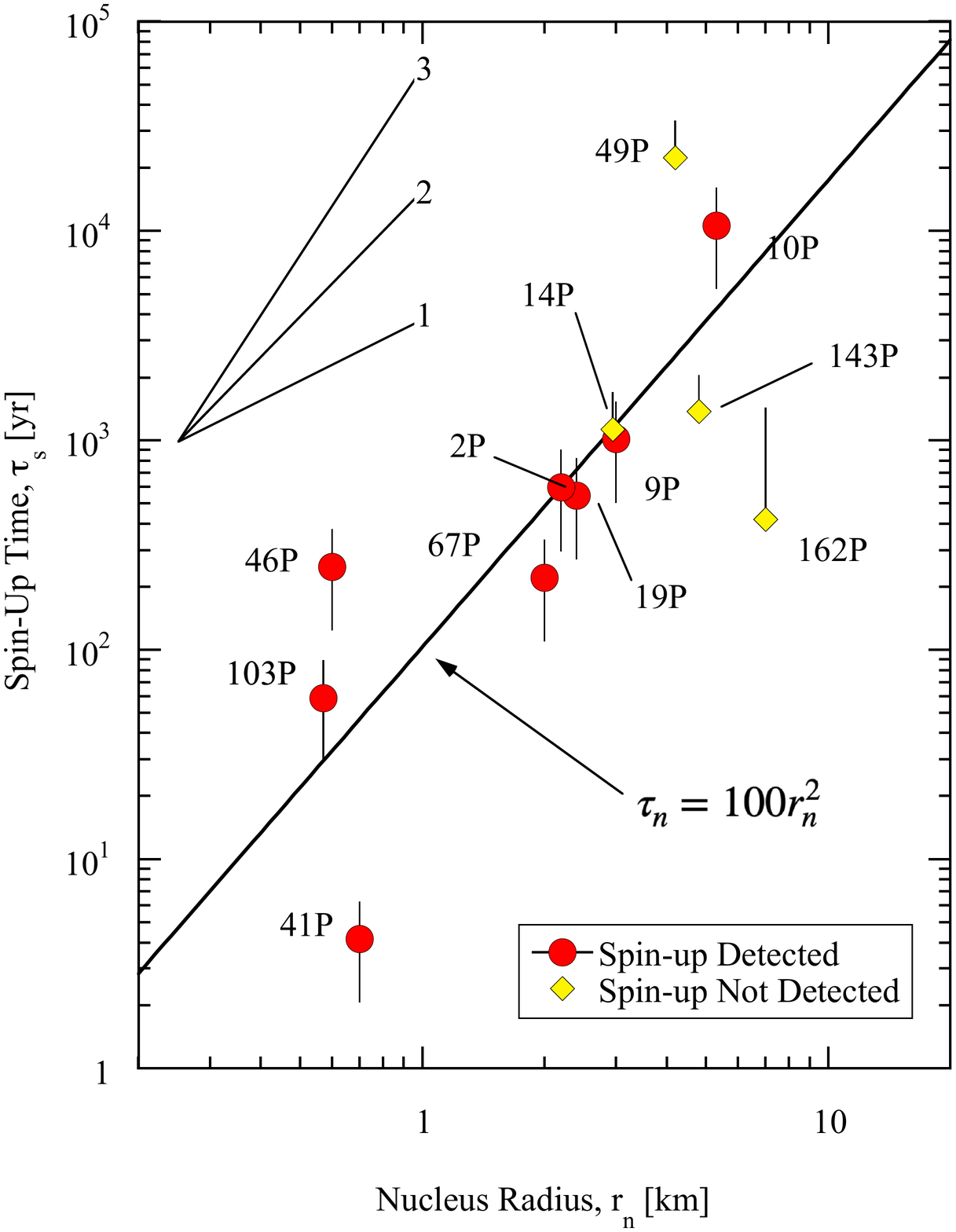}
\caption{Empirical spin-up timescale, $\tau_S$, vs.~nucleus radius, $r_n$, from Equation (\ref{timescale}) and Table (\ref{sublimation}). Filled red circles show comets in which period changes have been detected.  Filled yellow diamonds show comets in which only observational limits to period changes have been set.  Sample error bars show a $\pm$50\% uncertainty in $\tau_s$.  The solid line shows Equation (\ref{time}). Logarithmic slopes of 1, 2 and 3 are illustrated.  \label{tau_s_plot} }
\end{figure}

Figure (\ref{tau_s_plot}) shows $\tau_s$ as a function of nucleus radius, $r_n$, computed from Equation (\ref{timescale}) and the data of Table (\ref{sublimation}),  with illustrative error bars showing the effect of $\pm$50\% uncertainties in $\tau_s$.  Evidently,   $\tau_s$ varies widely in the range $\tau_s \sim$ 3 yr (for the very rapidly accelerating nucleus of 46P/Wirtanen) to $\tau_s \gtrsim 10^4$ yr (for 10P/Tempel 2 and 49P/Arend-Rigaux).     As a purely empirical diagram, the figure shows a  convincing, model-independent trend for larger values of $\tau_s$ to be  associated with  larger cometary nuclei, as expected based on scaling relations (Equation \ref{tau_s2}, Jewitt 1997) and noted by Samarasinha \& Mueller (2013) and Kokotanekova et al.~(2018).  It is obvious from the figure that $\tau_s$ and $r_n$ are related. Although numerical evidence of this is not needed, we computed the Spearman $\rho$ correlation coefficient (Press et al.~1992) between log($\tau_s$) and log($r_n$), finding $r_s$ = 0.88 and a $p$-value of 0.004, indicating a significant correlation.   A least-squares fit of a power law to the  comets having non-zero $|\Delta P|$ (red circles in  Figure \ref{tau_s_plot}) gives $\tau_s  = (102\pm50) r_n^{2.2\pm0.6}$.  However,  the significance of the fit should not be exaggerated (the sample is small, the uncertainties are poorly characterized, and we have ignored uncertainties in the radii of the comets plotted in Figure \ref{tau_s_plot}).  For convenience, we simply adopt 

\begin{equation}
\tau_s \sim 100 r_n^{2}
\label{time}
\end{equation}

\noindent with $\tau_s$ expressed in years and $r_n$  in kilometers, in the remainder of this paper, and point to Figure (\ref{tau_s_plot}) to show that this gives an acceptable match to the data.  It should be understood that this equation strictly applies to short-period comets with moderate eccentricities and perihelia near 1 AU (as indicated in Table \ref{sublimation}). Timescales for comets of a given size having different orbital semimajor axes and perihelia would not be fitted by Equation (\ref{time}).

\subsection{Active Fraction, $f_A$}
\label{AF}
Cometary mass loss is driven by the expansion of sublimated gas, the production rates of which are  estimated from the strengths of  resonance fluorescence bands, using a  model of the gas spatial distribution.  Typically, the Haser (1957) model, or one of its variants, is used to infer the production rate from spectroscopic data.  In most comets in the terrestrial planet region, the gas mass is dominated by sublimated water.  Accordingly, we use $\dot{M} = \mu m_H Q_{OH}$, where $\dot{M}$ is the mass production rate (kg s$^{-1}$), molecular weight $\mu$ = 18, and $Q_{H_2O}$ (s$^{-1}$)  is the production rate, most usually obtained from measures of the OH 3090\AA~(e.g.~A'Hearn et al.~1995) or Lyman $\alpha$ (Combi et al.~2019) bands.   Production rates can also be inferred from the strength of other gas species and even from the cometary continuum, but these are less reliable than OH production rates because of uncertainties in the relative abundances of species and of dust.  We do not use other species or dust measurements of production here.  We do note that, in many comets, the derived instantaneous dust mass production rates are larger than the gas mass production rates (i.e.~the ratio dust/gas $>$ 1).  Physically, though, dust speeds are small compared to the gas speed and the outflow momentum is necessarily dominated by the  gas. For these reasons, we here use only measurements of the gas production rates, and neglect the momentum carried by solids.

In order to examine $f_A(r_n)$ we combined active area determinations from the spectroscopic compilation by A'Hearn et al.~(1995) with Hubble Space Telescope-based nucleus radius measurements from Lamy et al.~(2004), to find 24 short-period comets common to both datasets (Table \ref{fAtable}). We added 126P/IRAS from Groussin et al.~(2004) to make a sample of 25.  The use of  two main sources reduces relative errors in the production rates introduced by different models and interpretations of the data.  Unavoidable systematic errors remain, however, notably from the unmeasured albedos and phase functions of the comets (however, infrared data examined by Fernandez et al.~(2013) suggest that albedo is not a strong function of nucleus radius).  For this reason, individual values of $f_A$ may differ somewhat from those reported by others in the literature.   As an example, consider 103P/Hartley.  We find (Table \ref{fAtable}) $f_A$ = 0.60, whereas Groussin et al.~(2004) reported $0.3 \lesssim f_A \lesssim$ 1 and Lisse et al.~(2009) reported $f_A$ = 1.1.  Values $f_A > 1$ are occasionally reported in some so-called ``hyper-active comets'', of which 103P/Hartley is one.  In such cases, the sublimation is presumed to come from grains in the coma, rather than from the nucleus directly.  Such grains cannot torque the nucleus.

\begin{figure}[ht]
\centering
\includegraphics[width=0.80\textwidth]{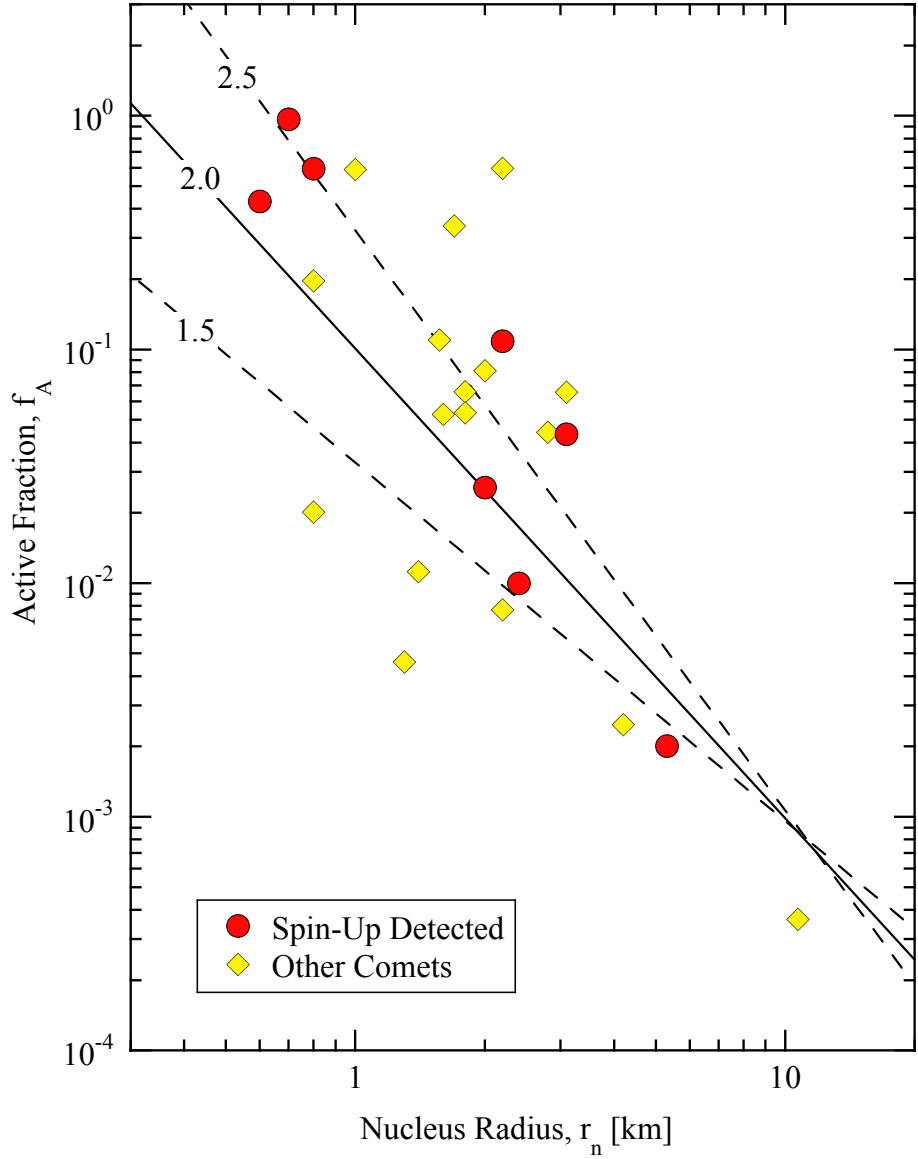}

\caption{Nucleus active fraction, $f_A$, as a function of nucleus radius, $r_n$.  The straight lines indicate $f_A \propto r_n^{-x}$ with x = 1.5,2.0,2.5 as marked.  \label{f_A} }
\end{figure}

The dependence of $f_A$  on $r_n$ for these comets is plotted in Figure (\ref{f_A}), where a strong inverse relation is evident.  The Spearman $\rho$  coefficient computed between log($f_A$) and log($r_n$) has value $r_s$ = -0.53 and a correspondingly low probability of being due to chance of $p$ = 0.008.    A least squares fit to all the data gives $f_A = 0.15\pm0.06 r_n^{-2.05\pm0.47}$.  A fit to the eight objects having measured spin changes gives $f_A = 0.22\pm0.08 r_n^{-2.61\pm0.52}$. The absolute uncertainties may be larger than indicated, and dominated  by systematic effects intrinsic to both the measurements and their interpretation.  For example, the ``Haser'' model gives a simplistic representation of the gas coma and production rates, with uncertainties which are both systematic and difficult to characterize.  The effective sublimating area is estimated through the adoption of a thermophysical  sublimation model whose parameters are themselves numerous and uncertain.  In addition, while there is no reason to remove the three largest nuclei from the plot, the effect of doing so would be to render $f_A$ independent of $r_n$, within the uncertainties\footnote{All  three of the largest nuclei have $f_A < 10^{-2}$, whereas this is true of only 2 of 22 (9\%) of the smaller nuclei. Assuming this same fraction, the chance of finding the three largest nuclei with $f_A < 10^{-2}$ is 0.09$^3 \sim 7\times10^{-4}$, which is significant at the $>$3$\sigma$ level of confidence.  }.  For all these reasons, and in order to avoid giving the appearance of undue significance to the relation in Figure (\ref{f_A}), we elect merely to note that the variation resembles the power law 

\begin{equation}
f_A \sim 0.1 r_n^{-2}
\label{fA}
\end{equation}

\noindent with $r_n$ in km. Figure (\ref{f_A}) shows that Equation (\ref{fA}) is a useful representation of the data.   
Equation (\ref{fA}) applies only for $f_A \le$ 1, which is true for   $r_n \gtrsim$ 0.3 km. Smaller nuclei should be entirely active ($f_A$ = 1)  over their surfaces by this relation.    

\subsection{Moment Arm, $k_T$}
We are interested to determine the dimensionless moment arm for the torque, $k_T$, since this quantity allows $\tau_s$ to be estimated through Equation (\ref{tau_s}) for any nucleus.  Substituting  Equation (\ref{timescale}) into (\ref{tau_s}) and solving for $k_T$ we find

\begin{equation}
k_T =  \frac{16\pi^2}{15} \left(\frac{ \rho_n r_n^4}{ V_{th} }\right) \left(\frac{ |\Delta P|}{P^2 P_K}\right)  \left(\frac{1}{\overline{\dot{M}}}\right)
\label{k_T}
\end{equation}

In Equation (\ref{k_T}), $r_n$,  $|\Delta P|$ and $P$  are measured quantities obtained from nucleus photometry and/or periodic coma structures that are modulated by nucleus rotation, while  $P_K$ is the orbital period.  The mean mass loss rate, $\overline{\dot{M}}$, is obtained from measurements of resonance fluorescence band strengths, focusing on the OH 3090\AA~band as a measure of the production rate of the dominant volatile, H$_2$O.  For practical reasons, measurements of  cometary spectra  tend to be taken near 1 to 2 AU.  Comets at distances $r_H \ll$ 1 AU appear at small elongations and are  difficult to measure.  Most comets at $r_H \gg$ 1 AU sublimate weakly and appear faint.   For these reasons, most of the spectroscopically well-observed comets (Table \ref{sublimation}) have perihelia $q \sim$ 1 to 2 AU.

The torque on the nucleus is maximized at perihelion where outgassing is strongest, but the total torque results from the total mass loss integrated around the orbit.   Accordingly, we apply a correction factor, $\mathcal{S}$, to  properly scale the mass loss rate measured at distance $r_H$, $\dot{M}(r_H)$, to estimate  the orbitally averaged mean mass loss rate, $\overline{\dot{M}}$, that would be measured if we possessed spectroscopic data around the orbit.  We define the scaling factor, $\mathcal{S}$ using

\begin{equation}
\mathcal{S}  = \frac{\overline{\dot{M}}} {\dot{M}(r_H)}
\label{bigS}
\end{equation}

\noindent where $\overline{\dot{M}}$ is the  mass loss rate averaged over one orbit period, $P_K$.  The calculation of $\mathcal{S}$ is described in the Appendix.   An obvious objection to the calculation of $\mathcal{S}$  is that the orbital variation of $\dot{M}$ might not be well represented by the sublimation model described in the Appendix.  For example, seasonal variations on comets with non-zero obliquity create important pre- vs.~post-perihelion asymmetries, but cannot be incorporated in the model.  We acknowledge this weakness and look forward to future gas production rate measurements taken more densely at a range of locations around the orbit.

For three of the comets in Table (\ref{sublimation}) (namely 14P/Wolf, 143P/Kowal-Mrkos, 162P/Siding Spring) we possess upper limits to the period change, $|\Delta P|$, but a literature search revealed no  measurements of the mass loss rates. Therefore, the moment arm for these three comets cannot be obtained from Equation (\ref{k_T}) and we exclude them from further consideration.  Conversely, while only an upper limit to the period change was set in 49P/Arend-Rigaux,  the mass loss rate has been quantified, and so we retain this plus seven other, better-measured comets in our sample to determine $k_T$ (Table \ref{sublimation}).  

Values of $\mathcal{S}$ are  listed for each nucleus in Table (\ref{sublimation}). 
The orbits of the well-characterized comets are  clustered near $q \sim$ 1 to 1.5 AU and $e \sim$ 0.6, for which  typical values are $\mathcal{S} \sim$ 0.1. This means that the sublimation rate averaged around the orbit of most comets is of order 10\% of the rate measured at perihelion.   2P/Encke has a smaller $q$ and larger $e$, resulting in $\mathcal{S} \sim$ 0.034.

We use  Equations (\ref{k_T}) and  (\ref{bigS}) to calculate $k_T$ for each nucleus.  The resulting values are listed in the penultimate column of Table (\ref{sublimation}).  
Values of the moment arm range from 2$\times10^{-4}$ to 4$\times10^{-2}$; the median value, $k_T$ = 0.007, 
is  an order of magnitude smaller than deduced (before observations) from an early toy model  ($k_T$ = 0.05, Jewitt 1997).  Our value for 9P/Tempel  ($k_T$ = 0.006) compares with the range 0.005$\le k_T \le$ 0.04 found by Belton et al.~(2011).  Our value for 103P/Hartley ($k_T$ = 4$\times10^{-4}$) is consistent with $k_T = 4\times10^{-4}$ from Drahus et al.~(2011).   

The data provide some evidence that $k_T$ and $r_n$ are correlated (Figure \ref{kt_vs_rn}).  The Spearman $\rho$ correlation coefficient between log($k_T$) and log($r_n$) is $r_s$ = 0.81, with a probability that this or a larger value could be obtained by chance of $p$ = 0.01.  The observed correlation is thus not statistically significant at the 3$\sigma$ ($p$ = 0.005) level.  A power law fit to the data gives $k_T = (1.4\pm0.8)\times 10^{-3} r_n^{1.6\pm0.5}$. The equation

\begin{equation}
k_T \sim 10^{-3} r_n^{2}
\label{k_t}
\end{equation}

\noindent adequately represents the data over the range 0.5 $\lesssim r_n \lesssim$ 7 km (Figure \ref{kt_vs_rn}).  The maximum possible value, $k_T = 1$, is reached at $r_n \sim$ 30 km, which is larger than any well-measured  cometary nucleus.


\subsection{Bias Effects}
We identify two sources of bias likely to affect determinations of $f_A(r_n)$.  First,  most comets are  discovered by magnitude-limited surveys, leading to a ``discovery bias''  acting against low activity (small $f_A$) comets  of a given size.  Small nuclei with small $f_A$ will be preferentially undercounted  relative to high activity (large $f_A$) comets of equal size because they are fainter.  The discovery bias is particularly acute for small nuclei, potentially pushing such objects with small active fractions beneath the survey detection threshold.  

Second, the determination of $f_A$ relies on spectroscopic measurements of  resonance fluorescence bands in gas.  These bands are weak in low activity comets of a given size, which therefore constitute more difficult and less appealing spectroscopic targets than bright comets (i.e.~those with large active areas). Small nuclei with small $f_A$, even if they survive the discovery bias, are thus likely to be under-reported in spectroscopic surveys  (e.g.~A'Hearn et al.~1995, Combi et al.~2018) of cometary activity, constituting a ``spectroscopy bias''.  

To examine this effect, we compiled the cumulative distribution of water production rates from the data listed by A'Hearn et al. (1995), the primary source of our activity data in Table (\ref{fAtable}).  The distribution shows a change in slope for $Q_{OH} \lesssim 10^{27}$ s$^{-1}$, corresponding to about $\dot{M}$ = 30 kg s$^{-1}$.   The local water ice sublimation rate at 1.5 AU is $f_s = 8\times10^{-5}$ kg m$^{-2}$ s$^{-1}$,  corresponding to a sublimating area $\dot{M}/f_s \sim$ 0.4 km$^2$, equal to the projected area of a circle of radius 0.35 km.  This is consistent with  the observation that small nuclei tend to be the most active, because sub-kilometer nuclei could not produce enough water to be spectroscopically detected in the A'Hearn survey if $f_A \ll$ 1.  

Quantitatively, Equations  (\ref{area}) and (\ref{fA}) show that the mass loss rate, $\dot{M} \propto f_A r_n^2$, is approximately independent of $r_n$, consistent with sublimation from a fixed active area (not fraction).  Substitution into these equations gives   $\dot{M} \sim 100$ kg s$^{-1}$, corresponding to $Q_{OH} \sim 3\times10^{27}$ s$^{-1}$ for a comet sublimating from the dayside hemisphere at representative distance $r_H$ = 1.5 AU (c.f.~Table \ref{sublimation}).    This is close to the limit of the spectroscopic data summarized in A'Hearn et al.~(1995), only 10\% of which have $Q_{OH} < 2\times10^{27}$ s$^{-1}$ ($\dot{M} \sim$ 60 kg s$^{-1}$).   Numerous, substantially less productive comets surely exist but are not spectroscopically attractive targets and are therefore under-reported.

A different bias likely plays a role in the distribution of the moment arm, $k_T$.   A small nucleus with a large $k_T$ would, by Equation (\ref{tau_s}), have a small spin-up time, leading to rotational instability and the removal of the nucleus from the observable population.  This ``survival bias'' results in an  observational sample that is naturally depleted of small nuclei having large values of the dimensionless moment arm, because these nuclei are less likely to survive (c.f.~Drahus et al.~2011).  The upper left portion of Figure (\ref{kt_vs_rn}) is presumably depleted of objects for this reason.  Conversely,  large nuclei, even if $k_T$ = 1, would take a long time to spin-up under the action of outgassing torques and so are less susceptible to the survival bias.

The existence of these bias effects does not eliminate the possibility that there are real size dependencies in $f_A$ and $k_T$.  For example, larger nuclei may be better able to retain refractory surface mantles than smaller nuclei because of their larger surface gravity, resulting in more complete blockage of the gas flow and depressing $f_A$  (Rickman et al.~1990).  Large nuclei are also more efficient in the recapture of slowly ejected material that could build a rubble mantle (Jewitt 2002).  Samarasinha and Mueller (2013) suggested that torques from multiple sources on highly active (large $f_A$) nuclei should more nearly cancel out than on weakly active (small $f_A$) nuclei.  This would lead to small nuclei having small $k_T$, as is suggested by Figure (\ref{kt_vs_rn}).  Unfortunately, we do not yet possess information sufficient to distinguish such effects from those due to the detection, spectroscopic and survival biases.

\section{CONSEQUENCES}

\subsection{Paucity of Small Nuclei}
Equations (\ref{tau_s}) and (\ref{time}) show that small nuclei are particularly susceptible to outgassing torques and therefore to potential rotational breakup (Jewitt 1992, 1997, Samarasinha 2007), consistent with the observed paucity of small nuclei (Fernandez et al.~2013, Bauer et al.~2017).   Indeed, the spins of nuclei smaller than a critical radius, $r_{c} \sim$ 0.1 to 0.3 km, can be substantially modified  within a few orbits.   The observed fragmentation of the small nucleus of 332P/Ikeya-Murakami (radius $r_n \le$ 0.28 km) is a particular example of a sub-kilometer nucleus likely suffering rotational instability.  Perhaps not coincidentally its rotation period, $P$ = 2 hour, is very short (Jewitt et al.~2016).  Rapid period changes observed in the small nuclei of comets 41P/Tuttle-Giacobini-Kresak (radius 0.7 km, Bodewits et al.~2018, Schleicher et al.~2019), 46P/Wirtanen (0.6 km, Farnham et al.~2021) and 103P/Hartley (0.6 km, Drahus et al.~2011) also indicate strong torques and incipient rotational instability.  

In addition to potential destruction by spin-up, a spherical nucleus of mass $M = 4\pi \rho_n r_n^3/3$  also experiences the loss of volatiles.  The true timescale for  devolatilization,
$\tau_{dv}$,  is an intractable function of the time-varying active fraction, $f_A$, and the dynamical evolution of the comet, with some evidence that these two are interconnected (Rickman et al.~1990).  A crude estimate may be obtained from
$\tau_{dv} \sim M/\overline{\dot{M}}$, with $\overline{\dot{M}}$ given by Equation (\ref{area}), giving

\begin{equation}
\tau_{dv} \sim \frac{\rho_n r_n}{3 f_A \overline{f_s}}.
\label{dv}
\end{equation}

\noindent We compare $\tau_{dv}$ with $\tau_s$ as a function of nucleus radius in Figure (\ref{lifetimes}), which updates Figure 2 from Jewitt (1997) to incorporate the new findings about the radius dependence of $f_A$ (Equation \ref{fA}) and $k_T$ (Equation \ref{k_t}).  We computed the orbitally-averaged $\overline{f_s}$ for hemispheric sublimation from comets having perihelion $q$ = 1.5 AU and eccentricity $e$ = 0.5, representative of those in Table (\ref{sublimation}), finding $\overline{f_s} = 2\times10^{-5}$ kg m$^{-2}$ s$^{-1}$.   We set $P$ = 15 hours, this being the median period from Table (\ref{sublimation}) and we assume $f_A = 0.1 r_n^{-2}$ for $r_n \ge$ 0.3 km (Equation \ref{fA}) and $f_A$ = 1 otherwise.   The resulting  sublimation lifetime from Equation (\ref{dv}) is shown in Figure (\ref{lifetimes}) as a solid red line.  The spin-up time is shown in blue for two assumptions about the radius dependence of $k_T$.  First, the dashed blue line shows $\tau_s(a)$, the timescale computed assuming that Equation (\ref{k_t}) holds for all $r_n$, even though we possess no constraining data for $r_n <$ 0.3 km.  Second, the dash-dot blue line shows $\tau_s(b)$ computed assuming that $k_T$ ``saturates'' to its value at $r_n$ = 0.3 km, namely $k_T = 10^{-4}$, for $r_n <$ 0.3 km, and otherwise follows Equation (\ref{k_t}).  These two assumptions reflect our lack of knowledge of the size dependence of the moment arm, but usefully demonstrate a range of possible behaviors.  Finally, the black curves in Figure (\ref{lifetimes}) show the  combined lifetimes, $\tau = (\tau_s^{-1} + \tau_{dv}^{-1})^{-1}$, with the  lower (yellow circles, $\tau_s(a)$) and upper (green diamonds, $\tau_s(b)$ branches reflecting the two models for $k_T(r_n)$ at $r_n <$ 0.3 km.   We emphasize that Figure (\ref{lifetimes}) is simplistic (real nuclei are not spherical, the bulk density is assumed, we have neglected seasonal effects, the model of equilibrium water ice sublimation is no doubt too simple) and is also specific to orbits with $q$ = 1.5 AU and $e$ = 0.5.  Timescales can be scaled from the figure to other orbits in inverse proportion to 
$\overline{f_s}$.  

Figure (\ref{lifetimes}) shows that the  spin-up timescales are shorter than the devolatilization timescale for all  comets with $r_n \gtrsim$ 0.1 km, regardless of which model for $k_T(r_n)$ is used.  This size range encompasses all cometary nuclei measured to-date, and shows the importance of spin-up.    Very few sub-kilometer nuclei are known, relative to power law extrapolations from larger sizes (e.g.~Meech et al.~2004), consistent with their rapid destruction.  For example, measurements of short-period comets in the 1 to 5 km radius range reveal a differential power law size distribution, $n(r_n) dr_n \propto r_n^{-3.3\pm0.3}dr_n$ (Bauer et al.~2017) while Fernandez et al.~(2013) found $n(r_n) dr_n \propto r_n^{-2.9\pm0.2}dr_n$.   If these power laws extrapolated to smaller radii, we should expect the number of nuclei with $r_n >$ 0.1 km radius to be $\sim$100 times the number with $r_n >$ 1 km.  Even given the observational bias against the detection of smaller objects, this seems unlikely to be true.  Crater counts in the Kuiper belt source region  reveal an impactor population with differential index $q = -1.7\pm0.3$ in the radius range $0.1 \lesssim r_n \lesssim$1 km (Singer et al.~2019).   Setting aside the question of how the source population could be flatter than the nucleus size distribution, $q$ = -1.7 would still give a population of $r_n >$ 0.1 km comets some 10$^{0.7}\sim$5 times larger than that of $r_n >$ 1 km comets, inconsistent with the data.  But, regardless of the size distribution of Kuiper belt objects, the strong size dependence of the lifetimes shown in Figure (\ref{lifetimes}) explains the paucity of small nuclei.

Figure (\ref{lifetimes}) also shows the median dynamical lifetime of short-period comets (Levison and Duncan 1997), marked by a long-dashed horizontal line at $\tau_{dyn} = 4\times10^5$ yr.  A dotted  black horizontal line shows, $\tau_L = 1.2\times10^4$ yr, the physical lifetime inferred by the same authors as necessary to match the inclination distribution of the comets.
We see that, while devolatilization of the larger nuclei is very slow (and may be impossible due to the formation of impermeable refractory surface mantles not accounted for here)   spin-up times are $\tau_s \lesssim \tau_L$ for all comets with $r_n \lesssim$ 10 km.   Almost all studied comets are smaller than 10 km in radius.    For example, of the 25 comets in Table (\ref{fAtable}), only one (28P/Neujmin) is larger than 10 km in radius.  Therefore,   the  spins of all measured comets are liable to have  evolved from their source-region values in response to outgassing torques.

\subsection{Long Nucleus Rotation Periods}

Figure \ref{periods} compares the rotation period distribution of cometary nuclei from Table (\ref{sublimation}) with that of small asteroids from Waszczak et al.~(2015).  For the latter, we selected asteroids with absolute magnitudes 13 $\le H \le 18$ in order to sample objects  similar in size to those of the comets. The median period of nuclei from Table (\ref{sublimation}) is $P_n$ = 15.0 hours (12 objects).  The median period of the 3883 small asteroids is $P_a$ = 6.35 hours.   The medians and the cumulative distributions of the periods are obviously inconsistent (Figure \ref{periods}), a conclusion buttressed by the K-S test, which gives the probability that the two distributions could be drawn by chance from the same parent as $<10^{-4}$.  We also compared the asteroid distribution with the list of comet rotation periods compiled by Kokotanekova et al.~(2017) with the same result; the K-S probability that the two distributions could be drawn from the same parent is $<10^{-4}$.  

The simplest explanation is that the median period difference reflects the role of density in setting the critical period for rotational instability.  In the absence of cohesive forces, the critical period, $P_C$, at which equatorial centripetal acceleration equals local gravity, varies with density as $P_C \propto \rho^{-1/2}$, and also depends on the body shape.  Periods in the ratio $P_n/P_a \sim$ 2.4:1 would indicate densities in the ratio $\rho_a/\rho_n \sim$ 5.8:1.   The nominal nucleus density is $\rho_n$ = 500 kg m$^{-3}$ (Groussin et al.~2019) while the average densities of C-type and S-type asteroids are reportedly  $\sim$1500 kg m$^{-3}$ and $\sim$3000 kg m$^{-3}$ (Hanus et al.~2017),  indicating ratios $\rho_a/\rho_n \sim$ 3 and 6 respectively.  Within the uncertainties on $P_n/P_a$, these expected and observed ratios are probably compatible.

However, other effects may also contribute to $P_n/P_a$.    Torques drive nucleus rotations equally towards shorter and longer values, but drive the median period towards longer values.  This is because nuclei torqued to periods shorter than $P_C$ should be destroyed, leaving a survivor distribution biased towards  longer-lived, longer period objects.  This effect is a function of nucleus size, with small nuclei more affected than large nuclei given the size dependence of $\tau_s$. While not detectable in the existing meagre observational sample, it should be sought in the future as more abundant and accurate data become available.   

Lastly, observational biases inherent in the methods of period determination play a potentially crucial role in Figure (\ref{periods}).  For example, rotational modulation of the photometry in active comets is limited by aperture averaging to periods longer than the aperture crossing time (Jewitt 1991), imposing a bias against short periods that is not present in photometry of asteroids and other point sources. A similar bias affects rotation periods determined from rotation-modulated coma structures (spiral arms and arcs, c.f.~Samarasinha and A'Hearn 1991) because short-period nuclei will produce tightly-wrapped spirals that are more difficult to resolve than open spirals from longer period nuclei. Disentangling these and other bias effects will be difficult. Ideally, we need a comet rotation sample based only on well-sampled bare-nucleus photometry in order to make an accurate comparison with the asteroids.

\subsection{Destruction of Sungrazing Comets}  

Thousands of small sungrazing (small perihelion) comets are known (Battams and Knight 2017).  Most  are  members of the so-called Kreutz group, with perihelia in the 0.01 to 0.02 AU range, and are thought to be products of the recent disruption of a larger precursor body (Sekanina and Chodas 2002, 2007).  Despite not impacting the photosphere (the radius of the Sun is $R_{\odot}$ = 0.005 AU), few Kreutz sungrazers survive  perihelion, and the same is true for members of the Kracht, Marsden and Meyer groups, which have similar or slightly larger perihelia.  Instead, observations indicate that the sungrazers are destroyed (or, more precisely, rendered invisible) before they reach peak solar insolation at perihelion.  For example, photometric measurements  of  three Kreutz comets show  peak  brightness   near $r_H \sim$ 12 $R_{\odot}$ ($\sim$0.06 AU), with subsequent fading on the way to perihelion (Knight 2010).   Could rotational disruption be responsible?

We can answer this question most directly for C/2005 S1, which is one of the best-observed Kreutz sungrazing comets.  This object lost sodium (presumably by desorption from minerals) at rate $\dot{M_{Na}}$ = 2 kg s$^{-1}$ when at $r_H$ = 12 $R_{\odot}$ (0.06 AU) (Knight 2010).  Sodium is merely the most readily observed optical species; others are surely present but undetected, and may carry more mass.   Therefore we conservatively  interpret $\dot{M_{Na}}$ as setting a  lower limit to the rate of loss of mass from C/2005 S1.   We assume  $\rho_n$ = 500 kg m$^{-3}$, $k_T$ = 0.007, $V_{th}$ = 10$^3$ m s$^{-1}$, $P$ = 5 hour and note that the estimated radii of most sungrazers fall in the range  $ 1\lesssim r_n \lesssim$ 50 m  (Knight 2010).  The radius of C/2005 S1 is estimated to be $\sim$10 m  but, to  be conservative and so to over-estimate $\tau_s$, we set $r_n$ = 50 m.  Then, Equation (\ref{tau_s}) gives the extraordinarily short characteristic time $\tau_s < 1.3\times10^5$ s (about 1.5 day).   This timescale  is only $\sim$1\% of the $\sim$month-long free-fall time from 1 AU to the Sun, providing ample opportunity for mass loss torques to spin-up and rotationally disrupt the nucleus, if it has a weak, comet-like structure.  Once the nucleus breaks up, the resulting components themselves are subject to fragmentation on even shorter timescales, resulting in the catastrophic destruction of the object (c.f.~Sekanina and Chodos 2002).  The peak brightness of C/2005 S1 occurred at $r_H \sim$ 14 R$_{\odot}$ (Figure 9 of Knight 2010) suggesting that this marks the point of fragmentation.   Since we assumed that the radius is at top end of the range given by Knight (2010), we can infer that rotational breakup  is an important destructive process for all smaller Kreutz comets.  

We cannot conclude that rotational break-up is the only destructive process, and many others of potential importance have been elucidated by Brown et al.~(2015).  For example, sungrazers that enter the Sun's Roche sphere (radius $\sim 2 R_{\odot}$ or $\sim$0.01 AU) could, if strengthless, be sheared apart by solar tides.   Sublimation of water ice, if present, is also very strong.   Using the approach from Section 4.1,  a 50 meter radius water ice body at $r_H$ = 0.06 AU would sublimate away on the timescale $\tau_{sub} \sim 1.8\times10^5$ s (a few days ).  Thus, if ice is present, devolatilization through sublimation can compete with spin-up at this size.

However, rotational disruption does not require the presence of ice in sungrazers, only of mass loss.  In fact, we are not aware of  direct evidence for ice in C/2005 S1 and there is little evidence for it in any other sungrazers.  For example,   the emission spectrum of C/(1965 S1) Ikeya-Seki at $r_H \sim$ 0.3 AU was dominated by metal lines (Na, Ca, Cr, Co, Mn, Fe, Ni, Cu, V) probably released by thermal desorption or sublimation of rocks (Slaughter 1969), possible because of the  high temperatures found near the Sun.  This raises the possibility that some  sungrazing comets are not comets at all, but asteroids (rocks) scattered in to orbits with small perihelia and disintegrating in the heat of the Sun.

\subsection{Main Belt Comets}
The rotations of small asteroids may be influenced by radiation (``YORP'') torques, with a timescale for spin-up approximately given by 

\begin{equation}
\tau_{Y} \sim \psi \left(\frac{r_n}{1~\textrm{km}}\right)^2 \left(\frac{r_H}{1~\textrm{AU}}\right)^2
\label{yorp}
\end{equation}

\noindent where $\psi = 1.3\times10^{13}$ s, $r_n$ is in kilometers and $r_H$ in AU (Jewitt et al.~2017).  By this equation, a 1 km asteroid in a circular orbit at 3 AU has $\tau_Y \sim$ 4 Myr.  The relation is very approximate because the YORP effect is sensitive to (mostly unknown) specific details of each asteroid, including the  shape, rotation vector and detailed thermophysical properties (Statler 2009); Equation (\ref{yorp}) is just a guide to the order of magnitude of the YORP timescale.   

Most asteroids have a refractory composition and sublimate negligibly under the Sun's radiation field.  However, a sub-population known as the ``active asteroids'' lose mass, generating comae and dust tails that are obvious in optical data (Jewitt 2012).  The causes of activity in these objects are many and varied, ranging from impact, to rotational instability, thermal fracture, desiccation stresses and the sublimation of near-surface ice (Hsieh and Jewitt 2006).  Active asteroids driven by ice sublimation are referred to as ``main belt comets''. 

If present, outgassing torques on the nuclei of main-belt comets will exceed the YORP torque when $\tau_s < \tau_Y$.  Combining Equations (\ref{tau_s}) and (\ref{yorp})  gives (c.f~Jewitt et al.~2017)

\begin{equation}
\dot{M_C} > \frac{16\times10^{12} \pi^2 \rho_n}{15 \psi V_{th} P k_T}\left(\frac{r_n}{1~\textrm{km}}\right)^2 \left(\frac{1~\textrm{AU}}{r_H}\right)^2
\end{equation}

\noindent for the critical mass loss rate above which the resulting torque exceeds that from YORP. Then, substituting $\rho_n$ = 1500 kg m$^{-3}$ (to take account of the larger density of asteroids),  representative asteroid period $P$ = 5 hours, $k_T$ = 0.007, we find that sublimation torques are dominant over YORP when

\begin{equation}
\dot{M_C} \gtrsim 0.02 \left(\frac{r_n}{1~\textrm{km}}\right)^2 \left(\frac{1~\textrm{AU}}{r_H}\right)^2
\label{dotm}
\end{equation}

\noindent with  $\dot{M_C}$ in kg s$^{-1}$.  For example, on a $r_n$ = 1 km body at $r_H$ = 2.5 AU, sustained mass loss rates as small as $\overline{M_C} = 3\times10^{-3}$ kg s$^{-1}$ could generate a torque larger than the YORP torque.  Such tiny mass loss rates fall  below the current spectroscopically detectable limits ($\dot{M} \sim$ 1 kg s$^{-1}$, Jewitt 2012) and therefore the existence of sublimation spin-up of asteroids cannot be directly tested.  Working against the influence of outgassing torques on icy asteroids is the observation that strong outgassing is highly time-variable, with main-belt comets spending a large fraction of the total time in an inactive or weakly active state (Hsieh and Jewitt 2006)

As a specific example, we consider the  disrupted outer-belt active asteroid P/2013 R3, whose precursor body broke into numerous $\sim$100 m scale pieces (Jewitt et al.~2017).  Sustained comet-like sublimation as small as (Equation \ref{dotm}) $\dot{M} \gtrsim 3\times10^{-5}$ kg s$^{-1}$  could, in principle, have driven this precursor to breakup on a timescale short compared to the YORP timescale.   Mass loss at such a low level would be completely unobservable using existing techniques.  Low albedo ice exposed at the subsolar point at $r_H$ = 3 AU sublimates in equilibrium with sunlight at the rate 2.8$\times10^{-5}$ kg m$^{-2}$ s$^{-1}$, meaning that a strategically located   ice patch of only $\sim$ 1 m$^2$ could generate a YORP-beating torque.  Temporarily larger rates of sublimation could have the same effect.  So, while we possess no evidence that P/2013 R3 was rotationally disrupted by sublimation torques, neither can we  reject  this possibility.  Hybrid schemes are also possible.  For instance, an initial breakup of a body triggered by impact or the YORP torque could expose previously buried water ice to the Sun, leading to sublimation and the rapid spin-up and disintegration of the fragments by outgassing torques.    Such hybrid schemes might be necessary to prevent the otherwise very rapid spin-up of ice-containing asteroids in the main-belt.


\section{DISCUSSION}

\begin{figure}[ht]
\centering
\includegraphics[width=0.80\textwidth]{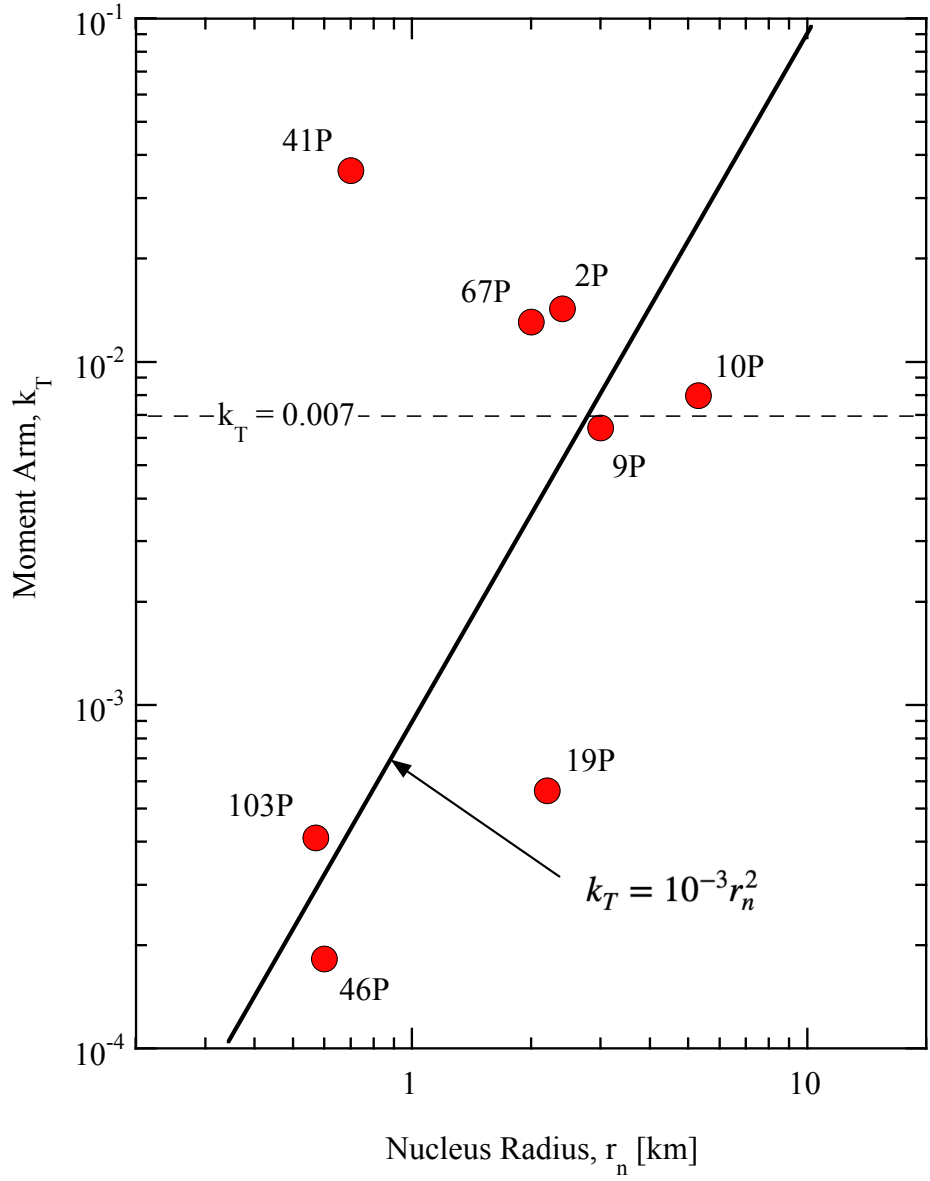}
\caption{Dimensionless moment arm, $k_T$, vs nucleus radius, $r_n$.    The median $k_T$ = 0.007 is marked by a dashed horizontal line and a fit is added to guide the eye. Data from Table (\ref{sublimation}).   \label{kt_vs_rn} }
\end{figure}

\begin{figure}[ht]
\centering
\includegraphics[width=0.70\textwidth]{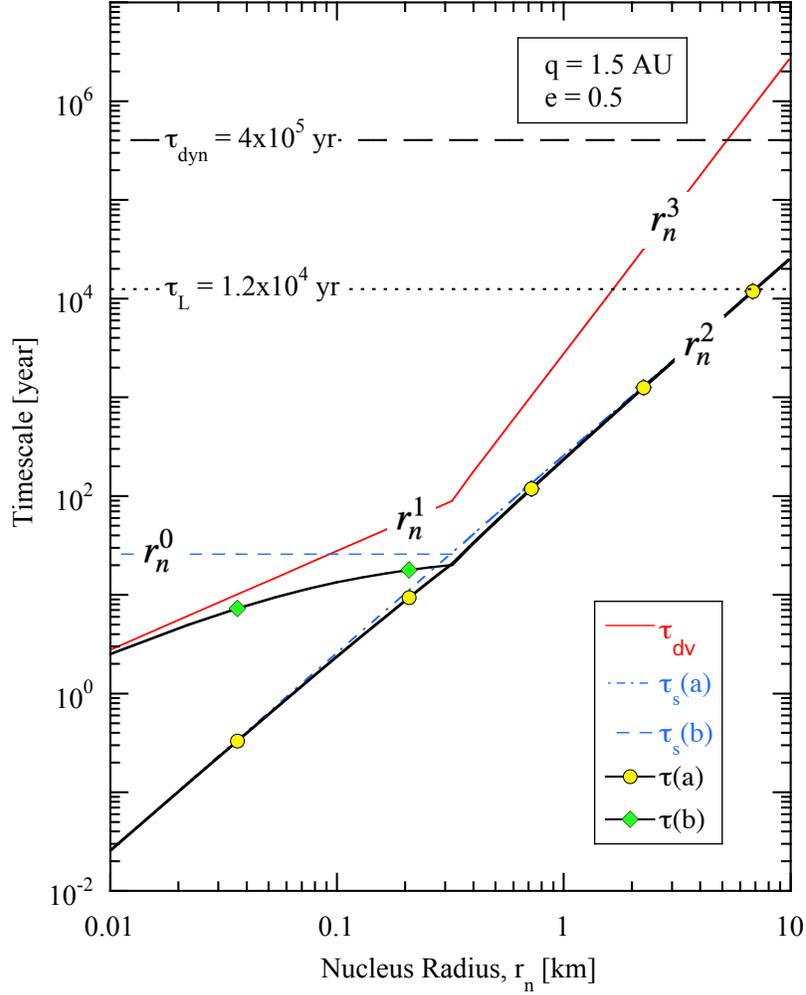}
\caption{Model lifetimes as a function of nucleus radius with respect to  spin-up ($\tau_s$, blue lines, from Equation \ref{tau_s2}) and devolatilization ($\tau_{dv}$, solid red line, from Equation \ref{dv}).  As discussed in the text, the dash-dot blue line, $\tau_s(a)$, assumes $k_T = 10^{-4}$ for $r_n \le$ 0.3 km and Equation (\ref{k_t}) otherwise. The dashed blue line, $\tau_s(b)$, assumes that Equation (\ref{k_t}) applies at all radii.    The combined lifetimes are shown as a thick black line with lower (yellow circles) and upper (green diamonds) branches labeled  $\tau(a)$ and $\tau(b)$, corresponding to the two models for $k_T(r_n)$.  Horizontal dashed and dotted black lines show the dynamical lifetimes of Jupiter family comets and their estimated active lifetimes, respectively, from a model by Levison and Duncan (1997). \label{lifetimes} }
\end{figure}

\begin{figure}[ht]
\centering
\includegraphics[width=0.80\textwidth]{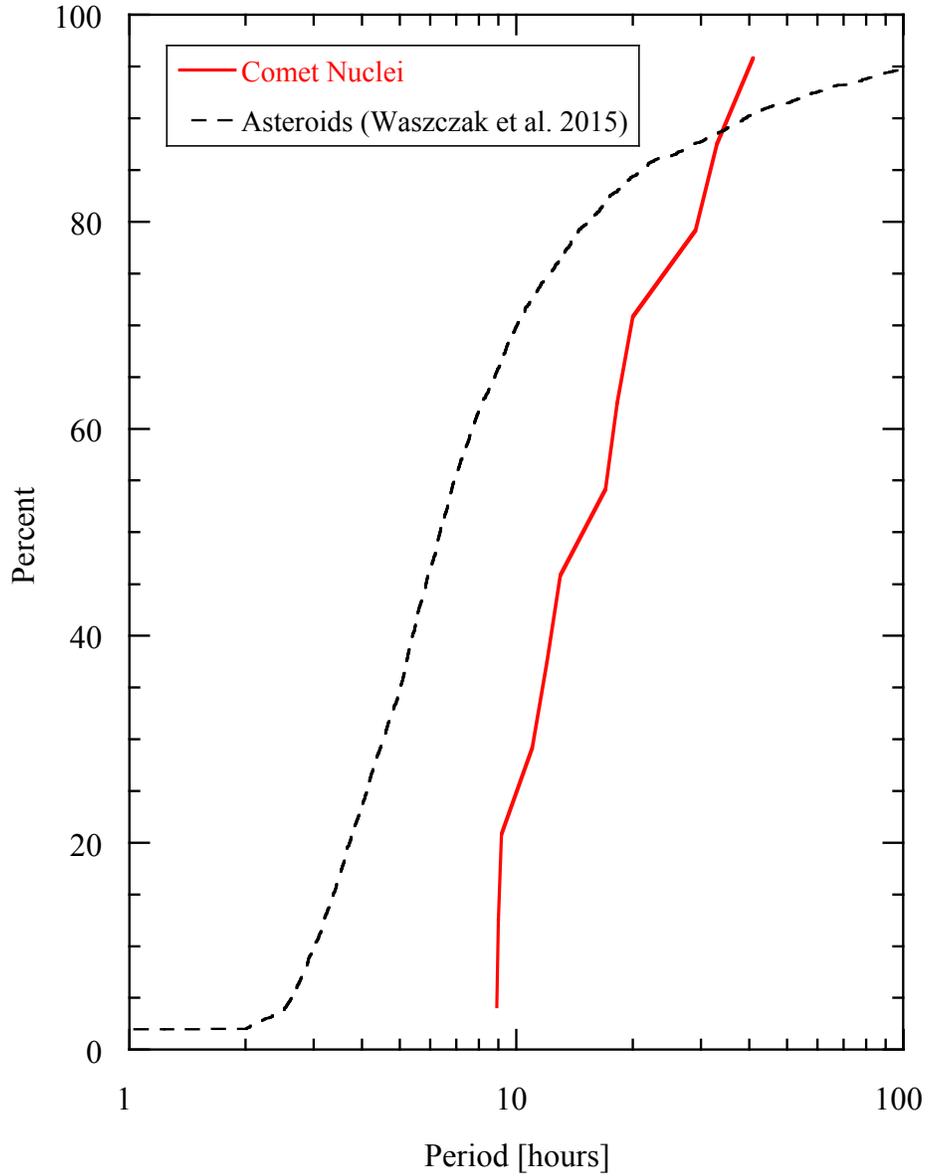}
\caption{Cumulative distributions of the rotation periods listed for (solid red line) comet nuclei in Table (\ref{sublimation}) and for (dashed black line) small asteroids from the sample of Waszczak et al.~(2015).   The eye and the K-S test confirm that these distributions are not consistent.  \label{periods} }
\end{figure}

\begin{figure}[ht]
\centering
\includegraphics[width=0.7\textwidth]{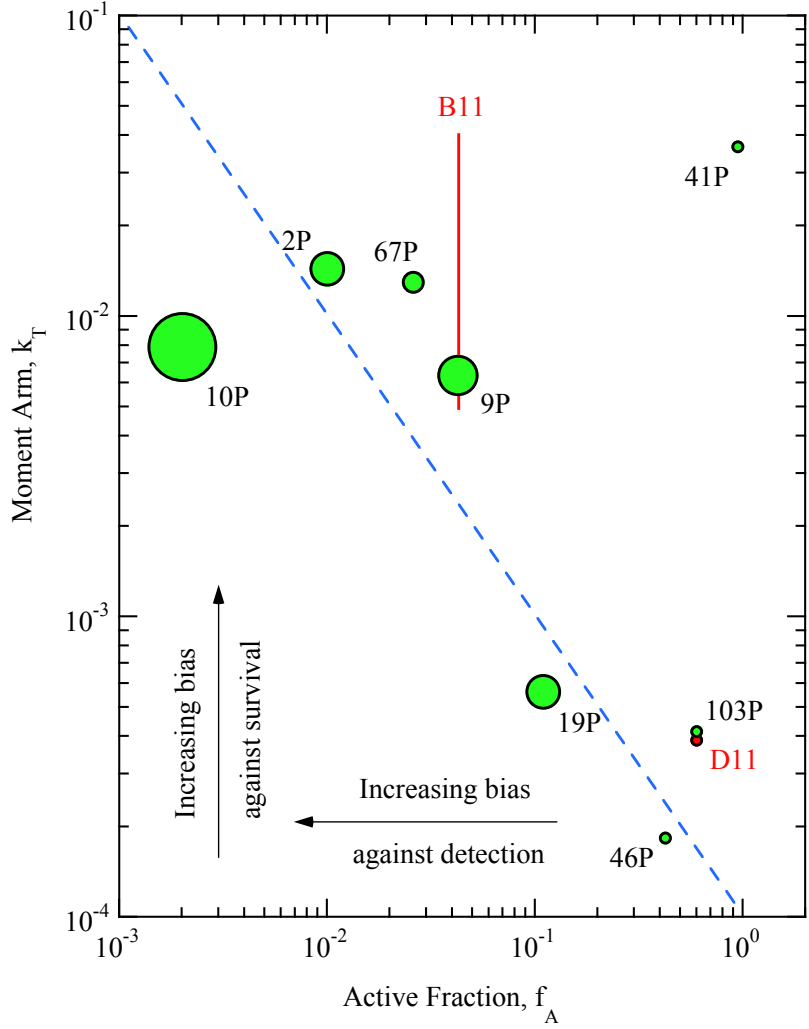}
\caption{Dimensionless moment arm, $k_T$, vs active fraction, $f_A$.    The  diameters of the symbols are proportional to the nucleus diameters.  Arrows show the direction of increasing discovery bias (which acts against small, weakly active comets because of their faintness) and increasing survival bias (which acts against small comets with large $k_T$ because of their vulnerability to rotational breakup, c.f.~Equation \ref{tau_s}).  The red line and circle marked B11 and D11 show, for comparison, values for 9P/Tempel by Belton et al.~(2011) and 103P/Hartley by Drahus et al.~(2011), respectively.  The dashed blue line indicates $f_A k_T = 10^{-4}$ and is not a fit to the data. 
\label{kt_vs_fA} }
\end{figure}

The observations establish beyond reasonable doubt both the importance of the outgassing torque in comets and the  major role played by observational selection effects.   To further emphasize these points, we refer to Figure (\ref{kt_vs_fA}), which shows  the moment arm, $k_T$, plotted against the active fraction, $f_A$, with the sizes of the plot symbols  shown in proportion to the radii of the nuclei.  Figure (\ref{kt_vs_fA}) illustrates three points.  First, the  discovery  bias against the detection of small cometary nuclei is evident from the top-heavy distribution of nucleus sizes.  Sub-kilometer nuclei are under-counted in optical surveys relative to their intrinsic proportion in the comet size distribution.  Second is the additional  bias against small comets with small active fractions, $f_A$, because for a given nucleus radius  the coma production rate (and hence the coma brightness and detectability) scales in proportion to $f_A$.  Small, weakly active nuclei are  pushed beneath the survey sensitivity limits leaving only small nuclei with large $f_A$, like 41P/Tuttle-Giacobini-Kresak, 46P/Wirtanen and 103P/Hartley (all with $f_A > 0.3$) in the figure.   Third,  is the survival bias against small nuclei having large $k_T$;  such objects have short spin-up times leading them to  be depleted in number by rapid breakup.  The notable outlier to this trend is 41P/Tuttle-Giacobini-Kresak, which is a small nucleus with a large moment arm and an empirical spin-up time that is exceedingly short (Figures \ref{tau_s_plot} and \ref{kt_vs_rn}). Howell et al.~(2018) suggested that 41P might be in an excited rotational state which, if true, would invalidate its inclusion  and improve the correlation with the remaining comets in Figures (\ref{kt_vs_rn}) and (\ref{kt_vs_fA}). At the other end of the  scale, the massive nucleus of 10P/Tempel 2 can sustain a large $k_T$  while still having a very long spin-up time.  The absence of large nuclei with small $k_T$ (lower left in Figure \ref{kt_vs_fA}) cannot be attributed to observational or survival bias.

 The dashed blue line in Figure (\ref{kt_vs_fA}) shows the relation $f_A k_T = 10^{-4}$, which evidently describes the observations  (with the exception of 41P/Tuttle-Giacobini-Kresak) rather well.   Re-arranging Equation (\ref{k_T}) and substituting Equation (\ref{area}) for $\overline{\dot{M}}$ we obtain
 
 \begin{equation}
k_T f_A =  \frac{4\pi \rho_n}{15 V_{th} P_K}     \left(\frac{|\Delta P| r_n^2}{P^2 \overline{f_s}}\right)
\label{k_T_2}
\end{equation}

\noindent where the orbitally-averaged sublimation flux, $\overline{f_s}$, is calculated as described in the Appendix.

Given that $\rho_n$, $V_{th}$ and $P_K$ are approximately the same for all comets in the study, the inference to be drawn from Equation (\ref{k_T_2}) and Figure (\ref{kt_vs_fA})  is  that $|\Delta P| r_n^2/(P^2 \overline{f_s})$ is constant.   This quantity, appearing in parentheses in Equation (\ref{k_T_2}), corresponds to the ``X parameter'' discussed by Samarasinha and Mueller (2013), Mueller and Samarasinha (2018) and Steckloff and Samarasinha (2018).    Steckloff and Samarasinha (2018) concluded that the near constancy of $X$ (but not for 41P/Tuttle-Giacobini-Kresak, as noted by Bodewits et al.~2018)  implied that  ``the net sublimative torque experienced by a comet nucleus depends predominantly on its size and heliocentric distance, independent of nucleus age, shape, local topography, and active fraction.''   Instead, Equations (\ref{T}) and (\ref{area}) show that the torque must depend on $f_A$, but that size-dependent trends in $f_A$ are largely canceled by those in the moment arm $k_T$, such that $k_T f_A \sim$  constant (c.f.~Figures (\ref{f_A}) and (\ref{kt_vs_rn})).  The approximate constancy of the $X$ parameter is  seen as a product of   these opposing size-dependent trends. 

Finally, a limitation of this and all investigations of nucleus rotation is the implicit assumption that the outgassing properties of each nucleus, including $f_A$ and $k_T$, remain fixed in time. In fact, the comets are dynamic and evolving bodies whose properties change both stochastically and in response to dynamical and thermal evolution.  As the surface and angular pattern of the mass loss evolve, the magnitude and possibly the direction of the sublimation torque might change.  Exactly this circumstance was reported in 46P/Wirtanen (Farnham et al.~2021), when the period change in the $\sim$50 days before perihelion was largely cancelled by the change after it.  Just as is the case with the YORP torque, whose magnitude and direction  change in response to even minimal disturbances of the surface (Statler 2009, Cotto-Figueroa et al.~2015), secular evolution of the sublimation torque vector can slow the rate of change of the nucleus angular momentum relative to the relations presented here.

\clearpage

\section{SUMMARY}
Anisotropic outgassing  exerts a torque which can change the spin of the cometary nucleus.  We parametrise the outgassing torque  in terms of the  radius of a spherical nucleus, $r_n$, the  fraction of the surface which is active, $f_A$, the dimensionless moment arm, $k_T$, the period, $P$, and the characteristic spin-up time, $\tau_s$.   Based on a simple model  we expect that $\tau_s \propto r_n^2/(f_A k_T P)$ (Equation \ref{tau_s2}). Using published rotational measurements of short-period comet nuclei with $0.5 \lesssim r_n \lesssim 7$  km and with perihelia $q \sim$ 1 to 2 AU, we find that 

\begin{enumerate}

\item The  empirical spin-up times follow $\tau_s \sim 100 r_n^{2}$, with $\tau_s$ in years and $r_n$ in kilometers.

\item The fractional active areas  vary  as $f_A \sim  0.1 r_n^{-2}$.

\item The  median dimensionless moment arm is $k_T$ = 0.007 with weak evidence for a size dependence $k_T \sim 10^{-3} r_n^2$.  

\end{enumerate}

Consequences of the short  timescales include 

\begin{enumerate}

\item Sub-kilometer  nuclei are rapidly destroyed, explaining their paucity relative to power-law extrapolations from larger sizes. This result is independent of the size distribution in the Kuiper belt source population.

\item The spin-up times of  sungrazing comets (most of which are small, $r_n \lesssim$ 50 m) are shorter even than the free-fall time to the Sun, consistent with their observed failure to survive passage through perihelion.

\item Weak mass loss torques on small main-belt asteroids, even at immeasurably small mass loss rates $\lesssim$ 1 g s$^{-1}$, surpass the YORP torque and, if sustained,  can control the spin-state.

\item The angular momenta of short period comets $\lesssim$10 km in radius are, on average, not primordial.

\end{enumerate}

Finally, we highlight  a)  flux-limited biases in optical and spectroscopic surveys against the discovery and measurement of nuclei with small $f_A$ and b) a survival bias against small nuclei with large moment arms, $k_T$, because these objects are quickly spun-up to rotational instability and removed from the observable population.  These importance of these biases should be assessed in future work.

\acknowledgments
I thank Jane Luu, Pedro Lacerda and the anonymous referee for helpful comments on this work.

\clearpage

\appendix{Appendix}
\setcounter{equation}{0}
\renewcommand{\theequation}{A\arabic{equation}}

In order to evaluate Equation (\ref{bigS}), we consider energy balance  for a sublimating surface, neglecting conduction, in the form

\begin{equation}
\frac{L_{\odot}(1-A)}{4\pi r_H^2} = \chi\left[\varepsilon \sigma T^4  + f_s(r_H) L(T)\right].
\label{energy}
\end{equation}

\noindent Here, $A$ and $\varepsilon$ are the Bond albedo and emissivity of the sublimating surface, $L_{\odot}$ is the solar luminosity, $r_H$ is heliocentric distance expressed in meters, $\sigma$ is the Stefan-Boltzmann constant and $L(T)$ is the temperature-dependent latent heat of sublimation.  We assume $A$ = 0.04, $\varepsilon$ = 1 while noting that solutions to Equation (\ref{energy}) are insensitive to both quantities.  Parameter $\chi$ is a dimensionless number that expresses the distribution of absorbed energy over the nucleus, varying between $\chi$ = 1 for a flat surface oriented perpendicular to the Sun-comet line and $\chi$ = 4 for an isothermal sphere.  We adopt $\chi$ = 2 as the intermediate case, corresponding to hemispheric warming of a spherical nucleus.   We solved Equation (\ref{energy}) using thermodynamic parameters for water ice tabulated by Brown and Ziegler (1980) and Washburn (1926).   The equilibrium temperature, $T$, was calculated as a function of $r_H$, which was in turn computed as a function of time by solving Kepler's equations

\begin{equation}
r_H(t) = a(1-e\cos(E(t))
\label{ellipse1}
\end{equation}

\begin{equation}
E(t) - e\sin(E(t)) = 2\pi (t - T_0)/P_K.
\label{ellipse2}
\end{equation}

\noindent Here, $E(t)$ is the eccentric anomaly and $T_0$ is the time of perihelion.   The specific sublimation rate $f_s$ was then used to evaluate $\overline{\dot{M}}$ from various combinations of $a$ and $e$ using Equations   (\ref{bigS}) and (\ref{area}).  The average sublimation rate is 

\begin{equation}
\overline{{f_s}} = \frac{1}{P_K} \int_0^{P_K} f_s(r_H) dt
\end{equation}

\noindent where the integral is taken around the orbit and, since $\dot{M} \propto f_s$, Equation (\ref{bigS}) becomes 

\begin{equation}
\mathcal{S}(r_H) = \frac{ \int_0^{P_K} f_s(r_H) dt}{P_K f_s(r_H)}.
\end{equation}




\clearpage


\clearpage

\begin{deluxetable}{lccccccccccccl}
\tabletypesize{\scriptsize}
\rotate

\tablecaption{Sublimation Spin-Up
\label{sublimation}}
\tablewidth{0pt}

\tablehead{Name & $a\tablenotemark{a}$ & $e\tablenotemark{b}$ & $q\tablenotemark{c}$ & $r_n\tablenotemark{d}$ & $P_K\tablenotemark{e}$   & $P\tablenotemark{f}$ & $|\Delta P|\tablenotemark{g}$ & $\dot{M}/r_H\tablenotemark{h}$ & $\mathcal{S}$\tablenotemark{i} & $\overline{\dot{M}}\tablenotemark{j}$ & $\tau_s\tablenotemark{k}$ & $10^3 k_T\tablenotemark{l}$ & Reference\tablenotemark{m}\\
& [AU] &   & [AU] & [km] & [yr]~ & [hr] & [min] & [kg s$^{-1}]$  &   & [kg s$^{-1}]$ & [yr]  &   \\}
\startdata

2P/Encke	&	2.215 &	0.848 &	0.337 & 2.4 & 3.30 & 11.0 & 4 & 1110/0.46 & 0.034 	& 38 & 540 		& 14 	& L04, K18, R18          \\
9P/Tempel	&	3.146 &	0.510 &	1.542 & 3.0 & 5.58 & 40.9 & 13.5 & 140/1.50 & 0.210	& 30 & 1014 	& 6 	&  L04, K18, G12          \\
10P/Tempel 	&	3.067 &	0.536 &	1.423 & 5.3 & 5.37 & 8.9 & 0.27 & 600/1.40 & 0.170 	& 102 & 10,600 	& 8  &  L04, K18, W17      \\
14P/Wolf	&	4.247	&	0.357	&	2.729	& 3.0 & 8.80 & 9.0	& $<$4.2 & --- & --- & --- & $>$1130 & --- 		&  F13, K18       \\
19P/Borrelly	&	3.611 &	0.624 &	1.358 & 2.2 & 6.86 & 29.0 & 20 & 1800/1.35 & 0.13 	& 596 & 600 		& 0.6  	&  L04, K18, M12       \\
41P/TGK	&	3.085 &	0.661 &	1.046 & 0.7 & 5.42 & 34.8	 & 1560 & 100/1.05 & 0.078 	& 8 & 4 		& 36  	&  L04, K18, C20          \\
46P/Wirtanen  	&	3.093 	&	0.659	&	1.055	& 0.6 & 5.44 & 9.15 & 12 & 390/1.06     & 0.078  & 30 & 250  	& 0.2 & L04, F21, C20  \\
49P/Arend-Rigaux	&	3.525 &0.619 &1.343 & 4.2 & 6.62 & 13.0 & $<$0.23 & 48/1.38 & 0.098 & 5 & $>$22,000 & $<$0.2 	& L04, K18, E17          \\
67P/C-G	&	3.465 & 	0.641 &	1.244 & 2.0 & 6.45 & 12.0 &  21 & 300/1.24 & 0.150 	& 45	 & 220 	& 13 	&  L04, K18, B19          \\
103P/Hartley	&	3.470 &	0.695 &	1.058 & 0.6 & 6.46 & 18.2 & 120 & 450/1.06 & 0.052 	& 23 & 60 		& 0.4 	&  A11, D13, C20            \\
143P/Kowal-Mrkos & 	4.298 &	0.409 &	2.542 & 4.8 & 8.90 & 17.0 & $<$6.6 & --- & --- & --- & $>$58.8 	& ---	& J03, K18 \\
162P/Siding-Spring &	3.050 &	0.596 &	1.232 & 7.0 & 5.30 & 33.0 & $<$25 & --- & --- & --- & $>$550 	& ---	& F13, K18\\

%

\enddata

\tablenotetext{a}{Orbital semimajor axis}
\tablenotetext{b}{Orbital eccentricity}
\tablenotetext{c}{Perihelion distance}
\tablenotetext{d}{Nucleus radius (Lamy et al.~(2004)}
\tablenotetext{e}{Orbital period}
\tablenotetext{f}{Rotation period (Kokotanekova et al.~(2018), except 46P from Farnham et al.~(2021)}
\tablenotetext{g}{Rotation change per orbit (Kokotanekova et al.~(2018), except 46P from Farnham et al.~(2021)}
\tablenotetext{h}{Reported mass loss rate (A'Hearn et al.~1995)}
\tablenotetext{i}{Scale factor, from Equation (\ref{bigS})}
\tablenotetext{j}{Orbit average mass loss rate, $\mathcal{S} \dot{M}$}
\tablenotetext{k}{Spin-up timescale, from Equation (\ref{timescale})}
\tablenotetext{l}{Dimensionless moment arm $\times10^3$, from Equation (\ref{k_T})}
\tablenotetext{m}{References: B19 = Biver et al.~2019, C20 = Combi et al.~2020, E17 = Eisner et al.~2017, G12 = Gicquel et al.~2012, J03 = Jewitt et al.~2003, K18 = Kokotanekova et al.~2018, L04 = Lamy et al.~2004, M12 = Macquet 2012, R18 = Roth et al.~2018, W17 = Wilson et al.~2017}

\end{deluxetable}

\clearpage
\begin{deluxetable}{lrrrcccl}
\tabletypesize{\scriptsize}
\tablecaption{Active Fraction Measurements
\label{fAtable}}
\tablewidth{0pt}

\tablehead{Name & $r_n\tablenotemark{a}$ & $A\tablenotemark{b}$ & $f_A\tablenotemark{c}$  
  \\}
\startdata
2P/Encke	  &  	2.4	  &  	0.7	  &  	0.010	\\
4P/Faye	  &   	1.8	  &   	2.7	  &   	0.066	\\
6P/d'Arrest	  &  	1.6	  &  	1.7	  &  	0.052	\\
9P/Tempel	  &   	3.1	  &   	5.2	  &   	0.043	\\
10P/Tempel	  &  	5.3	  &  	0.7	  &  	0.002	\\
19P/Borrelly	  &   	2.2	  &   	6.6	  &   	0.109	\\
21P/Giacobini-Zinner	  &  	1.0	  &  	7.4	  &  	0.590	\\
22P/Kopff	  &   	1.7	  &   	12.3	  &   	0.339	\\
26P/Grigg-Skjellerup	  &  	1.3	  &  	0.1	  &  	0.005	\\
28P/Neujmin	  &   	10.7	  &   	0.5	  &   	0.0004	\\
31P/Schwassmann-Wachmann	  &  	3.1	  &  	7.9	  &  	0.066	\\
41P/Tuttle-Giacobini-Kresak	  &   	0.7	  &   	6.0	  &   	0.970	\\
43P/Wolf-Harrington	  &  	1.8	  &  	2.2	  &  	0.054	\\
45P/HondaÐMrkosÐPajdusakova	  &   	0.8	  &   	0.2	  &   	0.020	\\
46P/Wirtanen	  &  	0.6	  &  	1.9	  &  	0.431	\\
47P/Ashbrook-Jackson	  &   	2.8	  &   	4.4	  &   	0.044	\\
49P/Arend-Rigaux	  &  	4.2	  &  	0.5	  &  	0.002	\\
59P/Kearns-Kwee	  &   	0.8	  &   	1.6	  &   	0.197	\\
67P/Churyumov-Gerasimenko	  &  	2.0	  &  	1.3	  &  	0.026	\\
68PKlemola	  &   	2.2	  &   	0.5	  &   	0.008	\\
74P/Smirnova-Chernykh	  &  	2.2	  &  	36.3	  &  	0.597	\\
78P/Gehrels	  &   	1.4	  &   	0.3	  &   	0.011	\\
81P/Wild	  &  	2.0	  &  	4.1	  &  	0.081	\\
103P/Hartley	  &   	0.8	  &   	4.8	  &   	0.595	\\
126P/IRAS	  & 	1.6	  &	3.4	  & 	0.110	\\					

%

\enddata

\tablenotetext{a}{Nucleus radius, [km], from Lamy et al.~(2004) except 126P/IRAS from Groussin et al.~(2004)}
\tablenotetext{b}{Active area, km$^2$, from A'Hearn et al.~(1995)  except 126P/IRAS from Groussin et al.~(2004) and 41P/TGK from Bodewits et al.~(2018)}
\tablenotetext{c}{Active fraction, $F_A = A/(4\pi r_n^2)$}

\end{deluxetable}



\clearpage 

\clearpage 

\clearpage

\clearpage

\clearpage

\clearpage

%

\clearpage


\end{document}